\documentclass[twocolumn]{aastex631}

\def\beq{\begin{equation}}
\def\eeq{\end{equation}}

\def\a{{\alpha}}
\def\b{{\beta}}
\def\D{{\Delta}}
\def\d{{\delta}}

\shorttitle{Evolution of Fast Radio Bursts}

\usepackage{graphicx}
\usepackage{enumitem}
\usepackage{float}
\usepackage{subfigure} 
\begin{document}

\title{Cosmological Evolution of Fast Radio Bursts and The Star Formation Rate}

\author{Sujay Champati}
\affiliation{California Institute of Technology, Pasadena, CA 91125, USA;
\underline{sujay@caltech.edu}}

\author{Vah\'e Petrosian}{1,2}
\affiliation{Department of Physics and KIPAC, Stanford University Stanford, CA 94305, USA; \underline{vahep@stanford.edu}}\\ 
\affiliation{Department of Applied Physics, Stanford University, Stanford, CA 94305, USA}

\begin{abstract}

We investigate the cosmological evolution of the luminosity and redshift of FRBs. As is the case for all extragalactic sources, we are dealing with  data that are truncated by observational selection effects, the most important being the flux limit, which introduces the so-called  Eddington-Malmquist bias.  In addition, for FRBs, there is a significant uncertainty in the redshifts obtained from the observed dispersion measures (DMs).  To correct for the truncation we use the non-parametric, non-binning Efron-Petrosian and Lynden-Bell methods, which give unbiased distributions of luminosities and redshifts and their cosmological evolution.  To quantify the redshift uncertainty, we use a data set which accounts for uncertainties of contribution to the DM of the host galaxy, and DM uncertainties due to fluctuation in the intergalactic medium. This data, in addition to a mean redshift, gives the one-sigma errors. We construct three  samples with  lower, mean, and upper redshifts and apply the above methods to each. For the three samples, we find similar (1) $\sim 3\sigma$ evidence for luminosity evolution, (2) a luminosity function that can be fit by a simple broken power law, and (3) a comoving density formation rate that decreases rapidly with redshift unlike the cosmic star formation rate (SFR). This rate is similar to that of short gamma-ray bursts, which are believed to result from compact star mergers with a formation rate delayed relative to the SFR. This may further support the hypothesis that magnetars are the progenitors of FRBs.
\end{abstract}


\section{Introduction} \label{sec:intro}

Since their discovery by \cite{Lorimer_2007}, fast radio bursts (FRBs), consisting of high-energy, short-duration radio pulses, have received considerable attention in about 20 monthly papers (see, e.g.~the FRB Newsletters)  addressing their progenitors,  energizing mechanisms, and emission processes. Their nearly isotropic distribution and large dispersion measures (DMs) indicate that most FRBs, except a few that are repeaters and/or are identified as galactic soft gamma-ray bursts,  have extragalactic origin. Several are also localized in host galaxies that provide their redshifts, $z$,  some with $z\sim 2$, showing a significant correlation with the measured DM. This has led to many papers (\cite{chen2024formationrateluminosityfunction}; \cite{James_2018}; \cite{zhang2024revisitingenergydistributionformation}; \cite{Lin_2024}; \cite{Zhang_2024}; \cite{Shin_2023}; \cite{Wang_2024}; \cite{zhang2022chimefastradioburst}) attempting to determine the cosmological evolution of their luminosities (or emitted energies), luminosity function, and, most importantly, evolution of their formation rate, which when compared with the cosmic star formation rate (SFR) can shed further light on their origin. However, such studies require  knowledge of the redshift of a large sample, ideally a sample with known observational selection effects. For sources with unknown host galaxy we can rely on the measured dispersion measures, based on the observed time delay as a function of frequency, $
\nu$; $\D t_{\rm obs}(\nu)\propto DM_{\rm obs}/\nu^2$. The observed $DM$ includes contribution from our galaxy, $DM_{\rm Gal}$, intergalactic medium, $DM_{\rm IGM}$, and the host galaxy, $DM_{\rm Host}$:
\beq
DM_{\rm obs}=DM_{\rm Gal}+DM_{\rm IGM}+DM_{\rm Host}/Z,
\label{DM}
\eeq
where $DM_{\rm Gal}$ includes contribution from the disk and the halo, and the host galaxy DM is divided by $Z\equiv(1+z)=\D t_{\rm obs}/\D t(z)$.%
\footnote{In what follows we use both $z$ and $Z=1+z$. The latter is more convenient and has more direct relation to time and distance than $z$.}
The contribution from IGM is equal to the column density of electrons along the line of sight, which depends on the baryon density and the cosmological model. Assuming a fully ionized IGM consisting of electrons, protons, and alpha particles, the co-moving electron density is $n_e=f_{\rm IGM}(\rho_b/m_p)(1-Y/2)$, where 
$\rho_b=\Omega_b(3H_0^2/8\pi G)$ is the co-moving baryon density, $f_{\rm IGM}$ is fraction of the baryons in the IGM, $Y$ is the fraction (by mass) of He, $\Omega_b\sim 0.05$, and $H_0=68$ km s$^{-1}$ Mpc$^{-1}$. Integrating the electron density along the line of sight to redshift Z we obtain 
\beq
\label{DMzRelation}
DM_{\rm IGM}(Z)=c\int_1^Z {Zn_e(Z)dZ/H(Z)},
\eeq
where $H(Z)=H_0\sqrt{\Omega_mZ^3+1-\Omega_m}$, for  the  $\Lambda$CDM model, with $\Omega_M=0.26$. In general $Y$ and $f_{\rm IGM}$ vary with redshift because of the production of He in stars and the accretion of  IGM gas  into galaxies and clusters. However, most of the post-primordial He production is locked in galaxies, and less than 10\% of cosmic baryons appear to be in galaxies and clusters. Thus, we can assume that these variables are constant $Y=0.25$, $f_{\rm IGM}\sim 0.9$, and we can use the above relations to estimate the redshift of FRBs without a known host. 

However, there are some uncertainties in the measured redshifts  due to two main factors. The first is the uncertainty of the non-IGM components of the DM, especially $DM_{\rm host}$, and the second is the uncertainty in $DM_{\rm IGM}$, arising from the cosmological evolution and inhomogeneities in the IGM. Quantitative accounts of these uncertainties are described in the next section.Nevertheless, as listed above, there have been many attempts to use thus obtained redshifts for the investigation of the cosmological evolution of FRBs.
Most of these use forward fitting (FF) methods that involve assuming parametric forms for several functions (luminosity function, luminosity and rate evolution functions, each with three or more parameters), often binning the data and ignoring the luminosity evolution, and  assuming  FRB formation rate the same as, or similar to, the SFR. We use the non-parametric, non-binning  method proposed by \cite{1992ApJ...399..345E}, which has proven useful in many areas, especially in the determination of the cosmological evolution of active galactic nuclei (AGNs) (\cite{Maloney_1999}; \cite{Petrosian_2014}; \cite{Singal_2022}; \cite{Zeng_2021}) and gamma ray bursts (GRBs) (\cite{Kocevski_2006}; \cite{Lloyd_1999}; \cite{2004ApJ...609..935Y}; \cite{Dainotti_2021}; \cite{Petrosian_2015}; \cite{Yu_2015}; \cite{Tsvetkova_2017}). Two of the papers cited above (\cite{chen2024formationrateluminosityfunction} and \cite{zhang2024revisitingenergydistributionformation}) also use this method. 

These methods are described in \S \ref{sec:methods}.  In the next section, we describe the FRB data we used. Another unique aspect of our work is that we include the $1\sigma$ range of the redshifts, with median, upper, and lower bound values. We used our method for each sample to obtain a range of cosmological evolutions. In \S \ref{sec:results} we show our results, and in \S \ref{sec:summary} we present a brief summary and conclusions.  

\section{CHIME Data} \label{sec:data}

The FRB catalog that we used in this analysis comes from the Canadian Hydrogen Intensity Mapping Experiment (CHIME). In a 1 year period between 2018 and 2019, CHIME detected 536 FRB events, including 62 repeat bursts from 18 repeaters, listed in CHIME Catalog 1 (\cite{2021}). 
This catalog gives the dispersion measure, $DM_{\rm obs}$, the integrated column density of electrons between the source and observer, which is used to obtain the redshift with Equations (\ref{DM}) and (\ref{DMzRelation}). This task requires the values of $DM_{\rm Gal}=DM_{\rm MW}+DM_{\rm Halo}$ and $DM_{\rm Host}$.

For $DM_{\rm halo}$, we use the approximation suggested by \cite{99b5141acd3547e8a7d84315ad40f119} of $50$ $\mathrm{pc~cm}^{-3}$. 
$DM_{\rm MW}$ has been studied by several authors, most notably \cite{cordes2003ne2001inewmodelgalactic} in the NE2001 models and \cite{Yao_2017} in the YMW2016 models. Both use the position (RA and Dec) of each source to determine the DM contribution of the Milky Way disk. 
To account for the uncertainty in the contribution of the host galaxy, several papers use  a log-normal probability distribution for $DM_{\rm host}$. For example, \cite{Tang_2023} obtain  the mean value $\langle \ln DM_{\rm host} \rangle = \ln (126.86)$ and dispersion $\sigma= 0.88$,  by fitting to a catalog of $11$ non repeating bursts with measured redshifts.
For the $DM_{\rm IGM}$, the common practice is to use the distribution  described by \cite{99b5141acd3547e8a7d84315ad40f119}, consisting of a Gaussian function (with a high value tail) for $\Delta = \rm DM_{IGM} / \langle  \rm DM_{IGM} \rangle$, as
\beq 
\label{DMIGMDistribution}
p_{\rm{IGM}}(\Delta)=A\Delta^{-\beta}\exp(-\frac{(\Delta^{-\alpha}-C_0)^2}{2\alpha^2\sigma_{\rm IGM}^2}), \, \Delta > 0, 
\eeq
where $\sigma_{\rm IGM}$ is the standard deviation. Hydrodynamic simulations by Macquart et al. also determine $\alpha = \beta = 3$, while $A$ and $C_0$ are chosen to normalize the distribution and set the mean value of $\Delta=1$.  

\begin{figure*}[]
    \centering
    \includegraphics[width=0.45\linewidth]{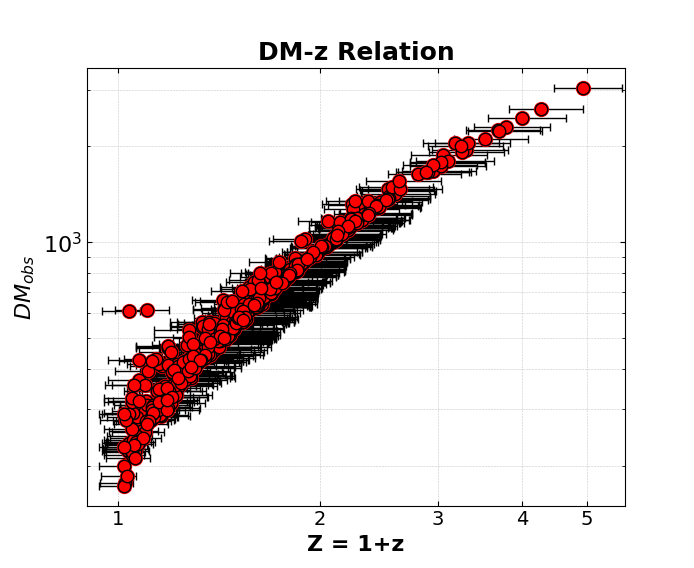}
    \includegraphics[width=0.45\linewidth]{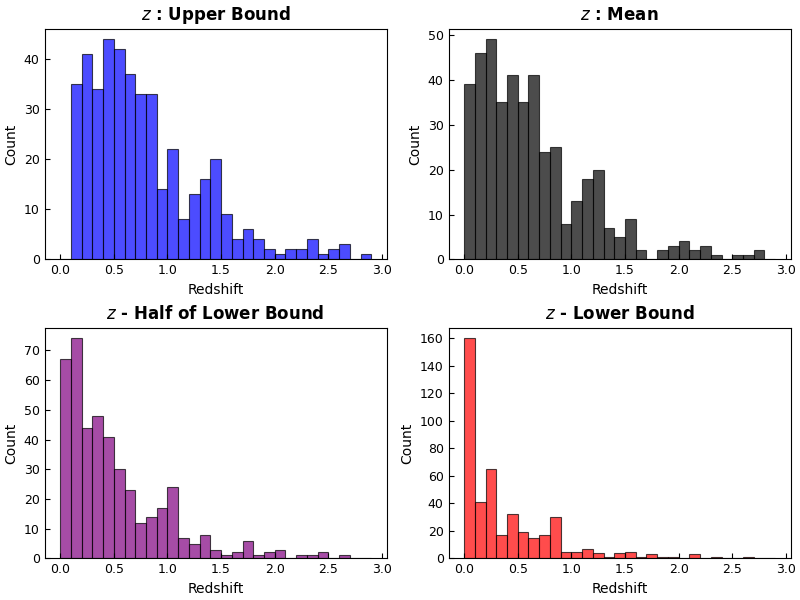}
    \caption{({\it Left}): Dispersion Measure - Redshift Relation. The points represent the mean values of redshift. The horizontal error bars show the possible lower and upper values. From \cite{Tang_2023}. ({\it Right}): Redshift histograms, from \cite{Tang_2023}, for mean-$z$ (top right), upper-$z$ (top left), lower-$z$ (bottom right), and half-lower-$z$ (bottom left). Note that some weaker $z<0.01$ are missing from the upper-$z$ sample.}
     \label{fig:dmzDistribution}
\end{figure*}

These uncertainties in $DM_{\rm obs}$,  translate into the uncertainties in $z$  shown in left panel of Figure \ref{fig:dmzDistribution}.  We take these uncertainties into account when performing our analysis of the evolution of the luminosity and the formation rate. 

In our data set, we first remove repeaters, since they are overrepresented in the catalog, and may possibly have different physical origins than the majority of the one-off sources. Furthermore, we do not use FRBs with $DM_{\rm obs} - DM_{\rm Gal} \leq 100 \rm{\frac{pc}{cm^3}}$ \footnote{$DM_{\rm obs} - DM_{\rm Gal}$ is often referred to as $DM_E$ in similar studies.} , as \cite{Tang_2023} do not compute redshifts for these bursts, on account of a high uncertainty in the dispersion due to the galactic foreground (MW+halo). This leaves us with $440$ remaining bursts in the raw sample.

Following the determination of the mean redshift of each FRB and its uncertainty, we show 
distributions of (1)  mean redshift $z$ (top right), (2) upper bound, $z_u=z+\d z_u$ (top left) and (3) lower bound, $z_l=z-\d z_l$ (bottom right) in the right panel of Figure \ref{fig:dmzDistribution}. We observe that the distribution of the lower bounds appears to be highly skewed and, as can be seen in the left panel of Figure \ref{fig:dmzDistribution},  a significant number of FRBs have negative $z$ or $\log Z$. To avoid this problem, we use a lower bound redshift with 50\% of the corresponding uncertainty, $z_l=z-(1/2)\d z_l$. The resultant distribution is shown in the lower left panel.

Given this fact, we choose to perform our analysis on the upper bound, mean, and $1/2$  of the lower bound (referred to as the lower bound below). We conduct our subsequent analyses on each distribution separately and combine them at the end to obtain the degree of uncertainty in our results.

Next, we computed the monochromatic peak luminosity, $L$, of each burst using the redshifts and the peak flux, $f(Z)$, measured by CHIME, as  
\vspace{-0.1cm}
\begin{equation}\label{LofZ}
        L (Z)= 4\pi  d_L^2(Z) \times f(Z)/{\bar K}(Z),
\end{equation}
with the luminosity distance $d_L(Z)=cZ\int_1^Z dZ/H(Z)$  and K-correction
$K(Z) = Z^{1+\a}$, where $\alpha=d\ln f(\nu)/d\ln \nu$, the  spectral index of fluxes, $f(\nu)$, estimated by \cite{Macquart_2019} to have an average value $\a = -1.5$.%
\footnote{Ideally one should use individual values of the spectral index for each source. Since the value of $\a$ is not known for all sources, we use the average value.}

We use the peak flux and luminosity instead of spectral energy and fluence used by some authors (e.g.~Zhang et al. 2025) because there is a redshift dependence bias in the latter quantities for transient sources (see e.g.~Kocevski \& Petrosian 2013: ApJ, 765:116), introduced by background radio noise: for higher Z more of the pulse tails fall below this noise, yielding shorter observed duration and smaller fluence. A recent study (\cite{frbcollaboration2024updatingchimefrbcatalogfast}) has shown that the catalog fluxes, based on intensity data, are lower limits due to uncertainty in source position within the beam. While more accurate fluxes are available for a subset of events in the CHIME Baseband Catalog, we retain the catalog values to maximize sample size and maintain consistency. We note that this may systematically underestimate luminosities, and future work could incorporate baseband-calibrated fluxes to refine the luminosity function and redshift distribution.
 
\begin{figure*}[]
    \centering
    \includegraphics[width=0.45\linewidth]{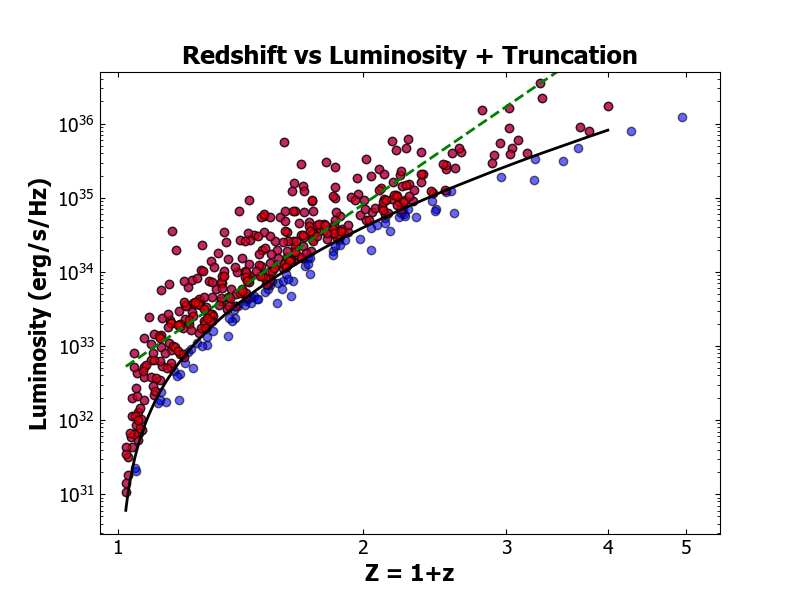}
    \includegraphics[width=0.45\linewidth]{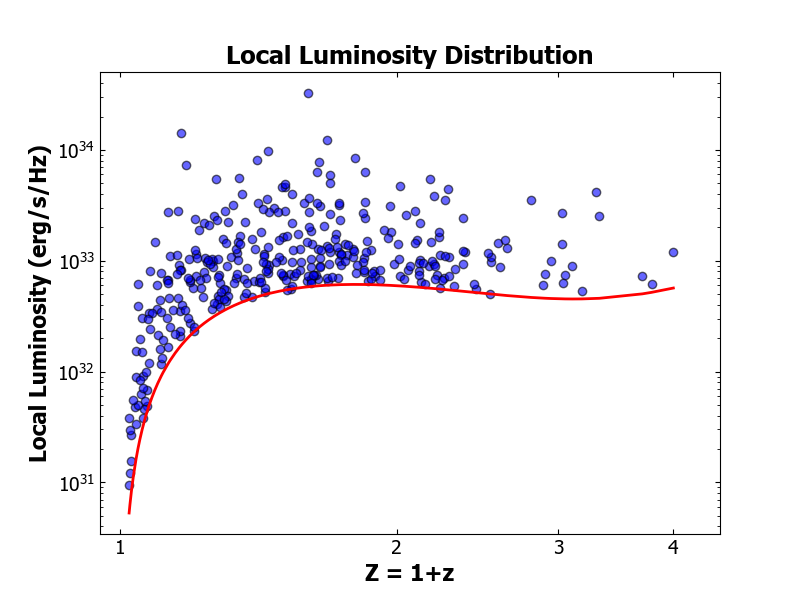}
        \caption{({\it Left}): Luminosity-redshift scatter diagram for the mean redshift sample with a linear regression fit (green dashed line) showing a strong luminosity evolution, $L\propto Z^{7.5}$, partly due to the truncation shown by the solid black curve based on Equation (\ref{Lmin}) with $f_{\rm lim}=0.5$ Jy. ({\it Right}): Scatter diagram of the now-independent variables local luminosity, $L_0$, and redshift, $Z$, with the de-evolved truncation boundary for the mean-$Z$ sample.}
       \label{fig:combined-LZ}
\end{figure*}

In the left panel of Figure \ref{fig:combined-LZ},  we show the scatter diagram of $L$ and $Z$ for the mean value sample. First, we note that there is a strong correlation between $L$ and $Z$; linear regression fit to the logarithmic values, shown by the green line, indicates a strong luminosity evolution, $L\propto Z^k$, with $k\sim 7.5$. Similar distributions and evolutions (with similar index $k$) are obtained for upper and lower $Z$ samples. Part of the correlation in the (raw) data is due to the truncation of the data caused by the flux limit, $f_{\rm lim}$. The nominal limit of the CHIME catalog is $f_{\rm lim}\sim 0.1$ Jy (\cite{2021}). Our analysis requires a ``complete" sample with robust truncation value. Thus, we choose a more conservative (i.e.~higher) flux limit, $f_{\rm lim}=0.5$ Jy, which eliminates a small fraction of sources, but assures completeness and minimizes  selection bias. A less robust flux limit (e.g.~those assumed by \cite{chen2024formationrateluminosityfunction}and \cite{zhang2024revisitingenergydistributionformation}) will lead to a stronger luminosity evolution and affect the rest of the analysis. From the flux limit we obtain  the truncation boundary
\beq
L_{\rm min}(Z) = 4\pi d^2_L(Z)f_{\rm lim}/{\bar K}(Z)
\label{Lmin}
\eeq
shown by the solid black line in the left panel of Figure \ref{fig:combined-LZ}. Thus, a source with $Z_i, L_i$ is truncated if $L_i<L_{\rm min}(Z_i)$ or if $Z_i>Z_{\rm max}(L_i)$ obtained from the inversion of the function: 
\beq
L_i = 4\pi d^2_L(Z_{\rm max})f_{\rm lim}/{\bar K}(Z_{\rm max}). 
\label{Zmax}
\eeq

\section{Methods and Approach} \label{sec:methods}

Our main goal is to determine the bivariate distribution $\Psi(L, Z)$, while accounting for the observational truncation described above, often referred to as the Malmquist or Eddington bias. As described in \cite{Petrosian1992}, most earlier and some recent attempts to  correct for this bias or truncation  use forward fitting methods, whereby one assumes parametric forms for the distributions and finds best fit value of the parameters. Later, nonparametric methods [e.g.~\cite{1968ApJ...151..393S} $V/V_{\rm max}$; \cite{1971MNRAS.155...95L} $C^-$]%
\footnote The Lynden-Bell method were later rediscovered by statisticians 
Woodroofe (1985, DOI: 10.1214/aos/1176346584) and Wang et al. (1986, www.jstor.org/stable/2241492). We refer to it as the $C^{-}$ method more familiar to the the astrophysics community.

were used to obtain the bivariate distributions directly from the data. However, as emphasized by \cite{Petrosian1992}, these methods assume that the variables in question are independent, i.e.~$\Psi(L, Z)=\psi(L)\rho(Z)$, which means that these methods ignore the possibility of luminosity evolution. To overcome this shortcoming,   \cite{1992ApJ...399..345E} (EP) developed a procedure that determines the correlation of variables in the presence of one-sided truncation.%
\footnote{In a subsequent paper, \cite{Efron01091999}, EP expand this procedure to two-sided truncated data.} 
We  use the EP method  to determine the luminosity evolution,  and after correcting for it, we define a new variable $L_0=L(Z)/g(Z)$ which is independent of $Z$, and determine the univariate distributions of the independent variables $L_0$ and $Z$ using the $C^-$ method.

These procedures have been used successfully in the investigation of cosmological distributions and evolutions of AGNs and GRBs in publications cited above, where more details of the procedures can be found.  
Briefly, the essence of the method involves constructing an associated set for each data point $(Z_i,L_i)$ based on the limiting luminosities $L_{min}(Z_i)$ and redshifts $Z_{\rm max}(L_i)$, and determining {\it the rank} of the data point by its luminosity $L_i$ or redshift $Z_i$ in the corresponding associated sets. The set of sources associated with $(Z_i,L_i)$ consists of all sources $(Z_j,L_j)$ with either $Z_j\leq Z_i$ and $L_j\geq L_{min}(Z_i)$, or $L_j\geq L_i$ and $Z_j\leq Z_{max}(L_i)$. 

Following the calculation of ranks, we determine the correlation using Kendall's Tau statistic, defined as
\begin{equation}
    \tau = \frac{\sum_i(R_i-E_i)}{\sqrt{\sum_i V_i }},
\end{equation}
where 
$R_i$ is the rank of the source $(L_i,Z_i)$ in its associated set, and  $E_i = \frac{N_i+1}{2}$ and $V_i = \frac{N_i^2-1}{12}$ are the expected mean and variance of the ranks, respectively, computed in terms of $N_i$, the number of points in the associated set of source i, so that for uniform distribution the expected normalized rank, $R_i/(N_i+1)$ is 1/2. A value of   $\tau =0 $ is expected for uncorrelated or independent variables. If  
$\tau$ is significantly different than zero its value gives the number of sigmas that the data  deviates from independence. We then use the new variable $L_0$, described above. For the evolution function $g(Z)$  we use a smooth broken power law form that flattens at high redshifts because of the decrease of the cosmic dynamic time at high redshifts, similar to ones used for GRBs (see, e.g.~\cite{Petrosian_2015}):

\begin{equation}
\label{e:LEfunction}
    g(Z) = Z^k \frac{1+Z_{cr}^k}{Z^k+Z_{cr}^k}
\end{equation}
Since $g(1)=1$, we refer to $L_0$ as the de-evolved {\it local} luminosity. The broken power law is chosen to prevent the rate of evolution to exceed the rate of the expansion, $H(Z)$, (or the inverse of the age). In the $\Lambda$CDM model, the expansion rate  flattens around $Z\sim 3$ to 4. Based on this and our past work on GRBs and AGNs (cited above, in particular \cite{Petrosian_2015}) we find that    $Z_{cr}\sim 3.5$ works well.  Independence is achieved for the value of $k$ for which $\tau=0$, and the $1\sigma$ range is given by the values of $k$ at 
$\tau=\pm 1$.

After determining the value of $k$ we use the $C^-$ method to calculate the {\it cumulative} local luminosity function 
\beq
\phi(L_0)=\int_{L_0}^\infty \psi(L')dL'
\label{cunLF}
\eeq
and the {\it cumulative} number rate 

\begin{equation}
    \dot \sigma(Z) = \int_1^Z \frac{\dot \rho (Z')}{Z'}\frac{dV(Z')}{dZ'}dZ',
\end{equation}
where ${\dot \rho (Z)}$ is the co-moving density formation rate and $V(Z)$ is the co-moving volume up to redshift $Z$.
For the cumulative luminosity function, we start by ordering the luminosities from highest to lowest, $L_1$ being the highest, and compute the cumulative function at $L_j$ using the following:

\begin{equation}
    \phi(L_j) = \phi(L_1)\prod_{i=2}^j (1+\frac{1}{N_i}),
\end{equation}
where $N_i$  is the number of sources in the associated set of source $i$, with $L_j\geq L_i$ and $Z_j\leq Z_{max}(L_i)$.$\phi(L_1)$ stands for the cumulative luminosity function above the highest observed luminosity $L_1$.

A similar procedure is used to compute the cumulative redshift distribution. However, here we start from the lowest redshift $Z_1$ and obtain:

\begin{equation}
{\dot\sigma}(Z_j)={\dot\sigma}(Z_1)\prod_{i=2}^j (1+\frac{1}{M_i}),
\end{equation}
where $M_i$  is the number of points in the associated set of source $i$,  with $Z_j\leq Z_i$ and $L_j\geq L_{min}(Z_i)$. Similarly, 
$\dot\sigma(Z_1)$ stands for the cumulative number rate between $Z=0$ and lowest  observed redshift $Z_1$.

\begin{figure*}[]
    \centering
   \includegraphics[width=0.47\linewidth]{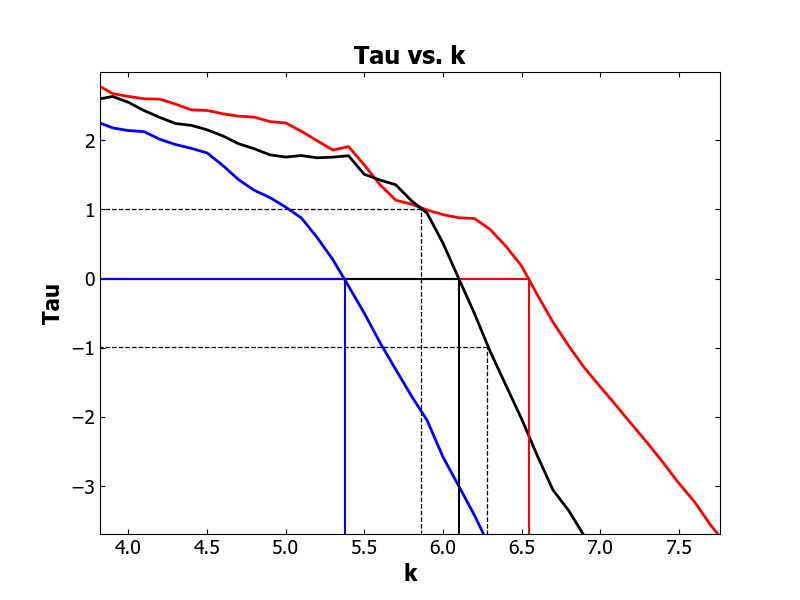}
   \includegraphics[width=0.45\linewidth]{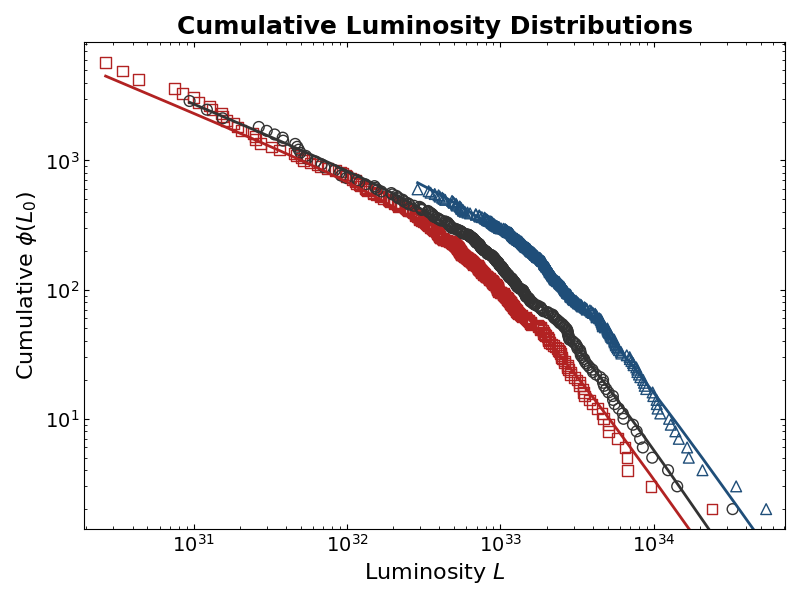}
\caption{({\it Left}:) Variation of Kendall's $\tau$ with the index $k$ of the luminosity evolution function (Eq.~\ref{e:LEfunction}), indicating best (shown by the three solid vertical lines) and one sigma values (shown by the dotted lines only for the mean-$Z$ sample); $k=5.3^{+0.3}_{-0.2}, 6.1^{+0.2}_{-0.3}, 6.5^{+0.2}_{-0.7}$ for upper (blue), mean (black) and lower (red) $Z$ samples, respectively. ({\it Right}:) Cumulative local luminosity function for the three samples (upper in blue, mean in black, and lower in red). The lines show the broken power fits based on Equation (\ref{cumLF}) with parameters given in Table \ref{tab:LFparams}.}
    \label{fig:combined_tau_cumLF}
\end{figure*}

\section{Results}
\label{sec:results}
 \subsection{Luminosity Evolution}
 \label{sec:LE}
 
The first task is to apply the EP procedure to the upper, mean, and lower $Z$ samples described in \S 2 and test for independence by computing the value of Kendall's $\tau$. We find that for the three samples $\tau(k=0)\geq 3.0$, indicating $\sim 3 \sigma$ evidence for luminosity evolution (often ignored). We then calculate the variation of $\tau$ with the index $k$ of the evolution function $g(Z)$, shown in the left panel of Figure \ref{fig:combined_tau_cumLF}, where we find that independence is obtained for $k\sim 5.3, 6.1$ and $6.5$ for the three samples, respectively. This allows us to calculate the local luminosity of each source. Figure \ref{fig:combined-LZ} (right) shows the local luminosities for the mean-$Z$ sample.

The next step is to calculate the mono-variate cumulative distributions $\phi(L_0)$ and ${\dot \sigma}(Z)$ using the Lynden-Bell $C^{-}$ method. 

\subsection{Luminosity Function} 
\label{sec:LF}

Figure \ref{fig:combined_tau_cumLF} (right) shows the cumulative (de-evolved) luminosity function for the three samples. As evident, we get similar shapes for the three samples. (Note that curve for upper-$Z$ sample ends at higher luminosity caused by the (un-physical) absence of FRBs with $z<0.1$.)  The three curves can be fit by a smoothly broken power law:

\beq\label{cumLF}
\phi(L_0)=\phi_0{(L_0/L_{\rm br})^{-\d_1}\over 1+ (L_0/L_{\rm br})^{\d_2-\d_1}},
\eeq
with slightly different fitting parameters: normalization, $\phi_0$, break luminosity, $L_{\rm br}$, and low (high) luminosity indices $\d_1  (\d_2)$ listed in Table \ref{tab:LFparams}.

\begin{table}[h!]
\centering
\caption{Luminosity Function Parameters}
\centering
\begin{tabular}{c|c|c|c|c}
     $\cdot$ & $L_{\rm br}$ (erg/s) & $\delta_1$ & $\delta_2$ & $\phi_0$ \\
     \hline
Lower Bound & $8.0 \times 10^{32}$    & $0.5$ & $1.7$   & $260$  \\
\hline
Mean        & $1.2 \times 10^{33}$   & $0.5$   & $1.75$  & $250$ \\
\hline
Upper Bound & $2.0 \times 10^{33}$  & $0.5$ & $1.7$    & $280$ \\
\hline
\end{tabular}
\label{tab:LFparams}
\end{table}
\subsection{Formation Rate  Evolution}
\label{sec:rate}

\begin{figure*}[]
    \centering
        \includegraphics[width=0.47\linewidth]{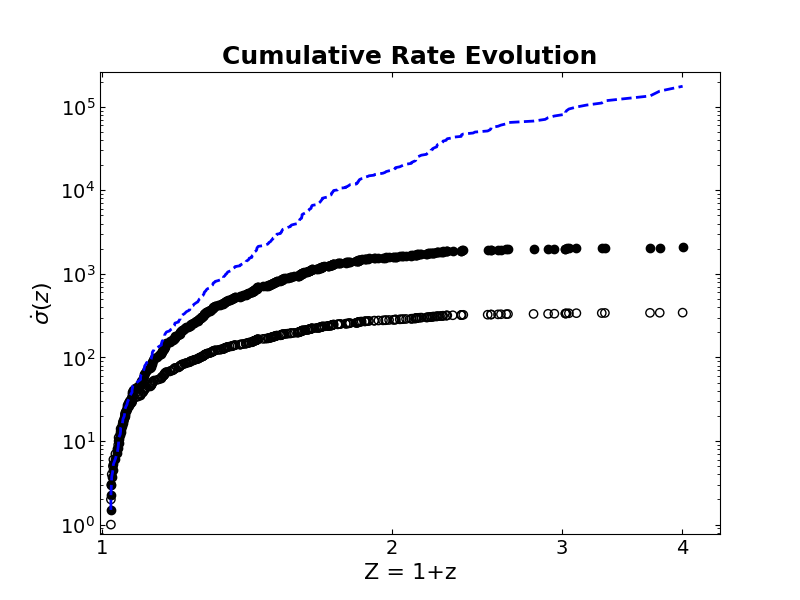}
        \includegraphics[width=0.43\linewidth]{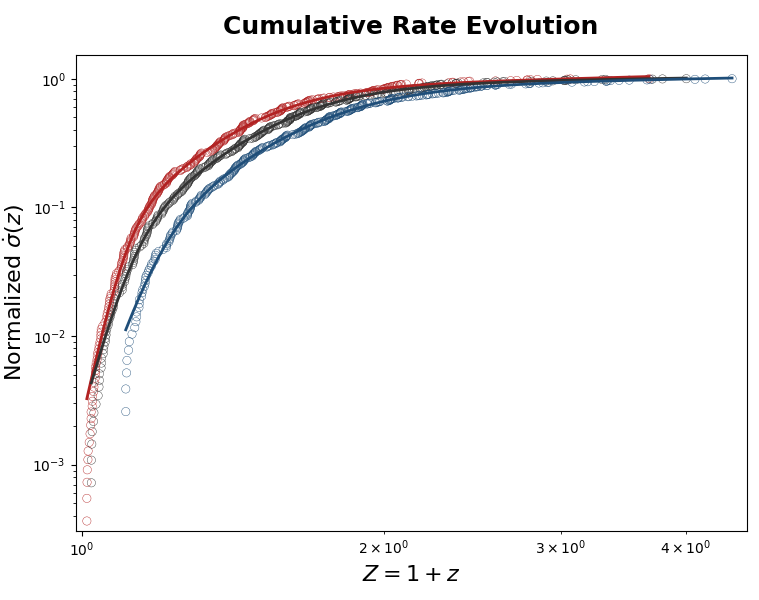} 
    \caption{({\it Left}): Cumulative formation rate numbers for the mean-$Z$ sample. The lower curve (open circles) shows the raw cumulative redshift counts, $N>Z)$. The middle curve is the correct number counts including correction for the luminosity evolution. The blue curve is for ignoring luminosity evolution.  ({\it Right}): Cumulative formation rate numbers for three samples (including luminosity evolution); red for lower, black for mean, and blue for upper, normalized at highest redshift. The curves show the fits obtained using Equation (\ref{densityrate}).}
       \label{fig:threesigmas}
   \end{figure*}

\begin{figure}
    \centering
    \includegraphics[width=1.0\linewidth]{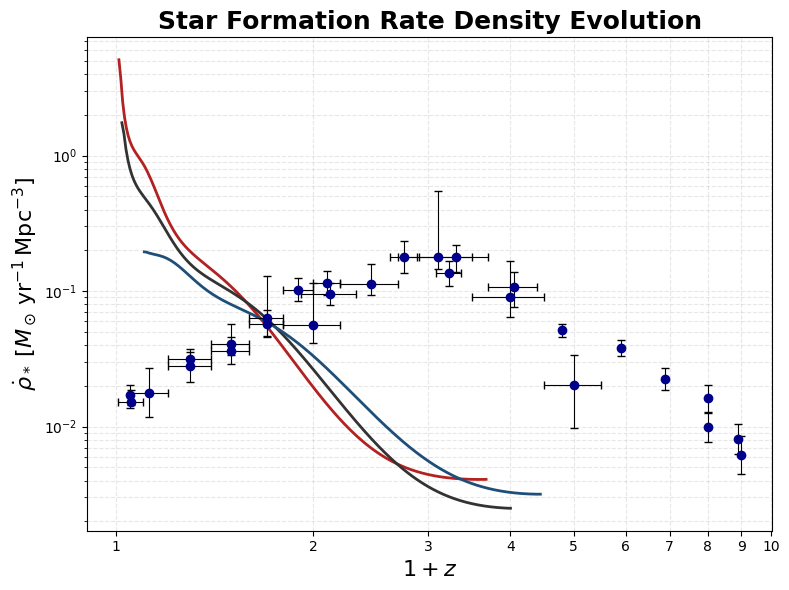}
    \caption{The Density Rate Evolution for three samples (red for lower, black for mean, and blue for upper, with arbitrary normalization) compared to the Star Formation Rate Taken From \cite{Madau_2014}. The unit on the y axis applies only to the SFR.}
    \label{fig:densitysfr}
\end{figure}

The left panel of Figure \ref{fig:threesigmas} shows the redshift variation of the cumulative number formation rate ${\dot \sigma}(Z)$ for the mean-$Z$ sample in full black points.  The black open points show the raw count, $N(>Z)$, representing the number of sources above $Z$. The upper blue curve is obtained ignoring the $\sim 3 \sigma$ evolution of the luminosity by applying the EP and $C^-$ methods to the data with the original truncation curve (Figure \ref{fig:combined-LZ} (left)). Figure \ref{fig:threesigmas} (right) shows the corrected (including luminosity evolution) formation rate numbers for the three samples (lower in red, mean in black and upper in blue). These are fit with broken power laws with two breaks
 \beq
 \label{fittosigma}
 {\dot \sigma}(Z)=\sigma_0Z^\a\times [1+(Z/Z_1)^{\a-\b}]^{-1}\times [1+(Z/Z_2)^{\b-\epsilon}]^{-1}
 \eeq
shown by the same color curves, and parameter values shown in Table \ref{tab:double_break_fit}. From  the derivatives $d{\dot \sigma}/dZ$, and  Equation (\ref{fig:densitysfr}),
we obtain  the co-moving density of the formation rate. 
\beq
\label{densityrate}
{\dot \rho}(Z)=Z{d\sigma(Z)/dZ\over dV(Z)/dZ}.
\eeq

The results are shown in Figure \ref{fig:densitysfr}, showing very similar shapes for the three samples, but distinctly different from the cosmic star formation rate (SFR).

\begin{table}[h]
\caption{Double broken power law fit coefficients for redshift distributions.}
\centering
\begin{tabular}{l c c c c c c}
\hline
 & $\sigma_0$ & $\a$ & $\b$ & $\epsilon$ & $Z_1$ & $Z_2$ \\
\hline
Lower-$Z$ & $0.002$ & $36.831$ & $7.820$ & $0.200$ & $1.105$ & $1.462$ \\
Mean-$Z$  & $0.003$ & $29.425$ & $7.118$ & $0.118$ & $1.124$ & $1.574$ \\
Upper-$Z$ & $0.001$ & $23.942$ & $6.484$ & $0.143$ & $1.190$ & $1.698$ \\
\hline
\end{tabular}
\label{tab:double_break_fit}
\end{table}


\section{Summary and Conclusions}
\label{sec:summary}

Our primary goal is the investigation of the cosmological evolution of the luminosity ($L$) and redshift ($Z=1+z$) distributions of FRBs, in particular the determination of their formation rate history compared with the cosmic SFR. There have been many attempts to this end, most of which used FF parametric methods, often assuming FRB formation rates similar to the SFR, (sometimes with small variation), and ignoring possible luminosity evolution. The unique feature of our work is the use of nonparametric, non-binning methods, whereby we first determine the luminosity evolution using the Efron-Petrosian procedure, which allows us to define a de-evolved (local) luminosity, $L_0$, that is independent of $Z$. This allows us to use the Lynden-Bell $C^{-}$ method to determine the univariate $L_0$ and $Z$ distributions. 

To our knowledge, two other papers (\cite{chen2024formationrateluminosityfunction}; \cite{zhang2024revisitingenergydistributionformation}) employ the Efron-Petrosian/Lynden-Bell method using a somewhat different sample and only the mean redshifts. They also use the CHIME nominal lower flux limit, resulting in a stronger luminosity evolution and a simpler evolution function $g(Z)=Z^k$, which yields a very high rate at $Z>3$. As mentioned at the outset, the greatest uncertainty in our work is the determination of the redshift from the observed dispersion measure $DM_{\rm obs}$ especially at low redshifts, where the contribution of the Milky Way  and the host galaxy  could be a large fraction of $DM_{\rm obs}$.  

In our work, we evaluate the effects of this uncertainty by defining three samples with redshifts equal to the mean, $1\sigma$ above the mean, called upper, and $0.5\sigma$ (instead of $1\sigma$, to avoid negative redshifts $z$) below the mean, called lower. We thus determine three sets of luminosity evolutions, luminosity functions, and co-moving density rate evolutions. 

Our main results are the following:

\begin{enumerate}
\item 
We find  $\sim 3 \sigma$ evidence for  luminosity evolution  and obtain relatively strong evolution rates  with indices $k=5.3, 6.1, 6.5$ for the upper, mean, and lower samples, respectively, (compared to GRBs with $k\sim 3$) defined in the evolution function, Equation (\ref{e:LEfunction}).

\item
Next we determine the cumulative luminosity functions, shown in Figure \ref{fig:combined_tau_cumLF} (right), which fit a broken power law, given in Equation (\ref{cumLF}), with very similar parameters (Table \ref{tab:LFparams}). Differentiation of this also results in  a smoothly broken power law (not shown) with high and low indices smaller by one (steeper). This is very similar to the luminosity function of many extragalactic sources, such as AGNs and GRBs.
\item 
Finally, we determine the evolution of the co-moving cumulative number rate, ${\dot \sigma}(Z)$, in Figure \ref{fig:threesigmas} (left), showing the raw counts and the formation rate with and without the luminosity evolution for the mean sample. In Figure \ref{fig:threesigmas} (right) we compare the ${\dot \sigma}(Z)$s of the three samples, showing very similar forms that are fitted 
to  power laws with two breaks coefficients shown in Table \ref{tab:double_break_fit}. Using the derivative of the fits in Equation (\ref{densityrate},) we calculate the evolution of the density rate, shown in Figure \ref{fig:densitysfr}. 
\end{enumerate}

Our main conclusion is that  the formation rate of FRBs is very different from the commonly assumed SFR, especially at low redshifts. This is very similar to the formation rate evolution found for short GRBs by \cite{Dainotti_2021}, whose progenitors are expected to be neutron star - neutron star or neutron star - black hole mergers and thus likely obey a formation rate delayed relative to the SFR and increasing at lower redshifts. As summarized in \cite{vahe2023progenitorslowredshiftgammaray} the formation rate of long GRBs also shows a rising low redshift component. Such a delayed formation rate would provide evidence to suggest that the progenitors of FRBs are magnetars.


\bibliographystyle{aasjournal}
\bibliography{main}

\end{document}